\documentclass[preprint]{aastex}

\usepackage{graphicx}
\usepackage{epstopdf}
\usepackage{amsmath}
\usepackage{algorithm,algorithmic}

\newcommand{\ie}{\textit{i.e.}, }
\newcommand{\eg}{\textit{e.g.}, }

\slugcomment{A title page comment if desired --- not to appear in the publication.}

\shorttitle{Astronomy in the Cloud: Using MapReduce for Image Coaddition}
\shortauthors{Wiley et al.}

\begin{document}


\title{Astronomy in the Cloud: Using MapReduce for Image Coaddition}


\author{K. Wiley, A. Connolly, J. Gardner and S. Krughoff}
\affil{Survey Science Group, Department of Astronomy, University of Washington}
\email{kbwiley@astro.washington.edu}

\and

\author{M. Balazinska, B. Howe, Y. Kwon and Y. Bu}
\affil{Database Group, Department of Computer Science, University of Washington}


\begin{abstract}
In the coming decade, astronomical surveys of the sky will generate tens of terabytes of images and detect hundreds of millions of sources every night.  The study of these sources will involve computation challenges such as anomaly detection and classification, and moving object tracking.  Since such studies benefit from the highest quality data, methods such as image coaddition, \ie astrometric registration, stacking, and mosaicing, will be a critical preprocessing step prior to scientific investigation.  With a requirement that these images be analyzed on a nightly basis to identify moving sources such as potentially hazardous asteroids or transient objects such as supernovae, these data streams present many computational challenges.  Given the quantity of data involved, the computational load of these problems can only be addressed by distributing the workload over a large number of nodes.  However, the high data throughput demanded by these applications may present scalability challenges for certain storage architectures.  One scalable data-processing method that has emerged in recent years is MapReduce, and in this paper we focus on its popular open-source implementation called \textit{Hadoop}.  In the Hadoop framework, the data is partitioned among storage attached directly to worker nodes, and the processing workload is scheduled in parallel on the nodes that contain the required input data.  A further motivation for using Hadoop is that it allows us to exploit cloud computing resources, \ie platforms where Hadoop is offered as a service.  We report on our experience implementing a scalable image-processing pipeline for the SDSS imaging database using Hadoop.  This multi-terabyte imaging dataset provides a good testbed for algorithm development since its scope and structure approximate future surveys.  First, we describe MapReduce and how we adapted image coaddition to the MapReduce framework.  Then we describe a number of optimizations to our basic approach and report experimental results comparing their performance.

\end{abstract}


\section{Introduction}
\label{secIntroduction}

Many topics within astronomy require data gathered at the limits of detection of current telescopes.  One such limit is that of the lowest detectable photon flux given the noise floor of the optics and camera.  This limit determines the level below which faint objects, (\eg galaxies, nebulae, and stars), cannot be detected via single exposures.  Consequently, increasing the signal-to-noise ratio (SNR) of such image data is a critical data-processing step.  Combining multiple brief images, \ie \textit{coaddition}, can alleviate this problem by increasing the dynamic range and getting better control over the point spread function (PSF).  Coaddition has successfully been used to increase SNR for the study of many topics in astronomy including gravitational lensing, the nature of dark matter, and the formation of the large structure of the universe.  For example, it was used to produce the Hubble Deep Field \citep{HDF}.

While image coaddition is a fairly straightforward process, it can be costly when performed on a large dataset, as is often the case with astronomical data (\eg tens of terabytes for recent surveys, tens of petabytes for future surveys.).  Seemingly simple routines such as coaddition can therefore be a time- and data-intensive procedure.  Consequently, distributing the workload in parallel is often necessary to reduce time-to-solution.  We wanted a parallelization strategy that minimized development time, was highly scalable, and which was able to leverage increasingly prevelant service-based resource delivery models, often referred to as \textit{cloud computing}.  The MapReduce framework (Google 2004)offers a simplified programming model for data-intensive tasks, and has been shown to achieve extreme scalability.   We focus on Yahoo's popular open-source MapReduce implementation called Hadoop (Hadoop 2007).  The advantage of using Hadoop is that it can be deployed on local commodity hardware but is also available via cloud resource providers (an example of the \textit{Platform as a Service} cloud computing archetype, or \textit{PaaS}).  Consequently, one can seamlessly transition one's application between small- and large-scale parallelization and from running locally to running in the computational cloud.

This paper discusses image coaddition in the cloud using Hadoop from the perspective of the \textit{Sloan Digital Sky Survey} (SDSS) with foreseeable applications to next generation sky surveys such as the \textit{Large Synoptic Survey Telescope} (LSST).

In section \ref{secBackgroundWork}, we present background and existing applications related to the work presented in this paper.
In section \ref{secMapReduce}, we describe cloud computing, MapReduce, and Hadoop.
In section \ref{secImplementingImageCoadditionMapReduce}, we describe how we adapted image coaddition to the MapReduce framework and present experimental results of various optimization techniques.
Finally, in sections \ref{secConclusions} and \ref{secFutureWork}, we discuss final conclusions of this paper and describe the directions in which we intend to advance this work in the future.

\section{Background and Related Work}
\label{secBackgroundWork}

\subsection{Image Coaddition}
\label{subsecImageCoaddition}

Image coaddition is a well-established process.  Consequently, many variants have been implemented in the past.  Examples include \textit{SWarp} \citep{SWarpA, SWarpB}, the SDSS coadds of Stripe 82 \citep{SDSSDR7}, and \textit{Montage} \citep{Montage} which runs on \textit{TeraGrid} \citep{TeraGrid}.  SWarp does support parallel coaddition; it is primarily intended for use on desktop computing environments and its parallelism is therefore in the form of multi-threaded processing.  Parallel processing on a far more massive scale is required for existing and future datasets.  Fermilab produced a coadded dataset from the SDSS database (the same database we processed in this research), but that project is now complete and does not necessarily represent a general-purpose tool that can be extended and reused.  Montage most closely related to our work but there are notable differences.  Montage is implemented using MPI (Message Passing Interface), the standard library for message-passing parallelization on distributed-memory platforms. MapReduce is a higher-level approach that offers potential savings in development time provided one's application can be projected well onto the MapReduce paradigm.  Hadoop in particular (in congruence with Google's original MapReduce implementation) was also designed to achieve high scalability on cheap commodity hardware rather requiring parallel machines with high-performance network and storage.

\begin{figure}
\epsscale{1}
\plotone{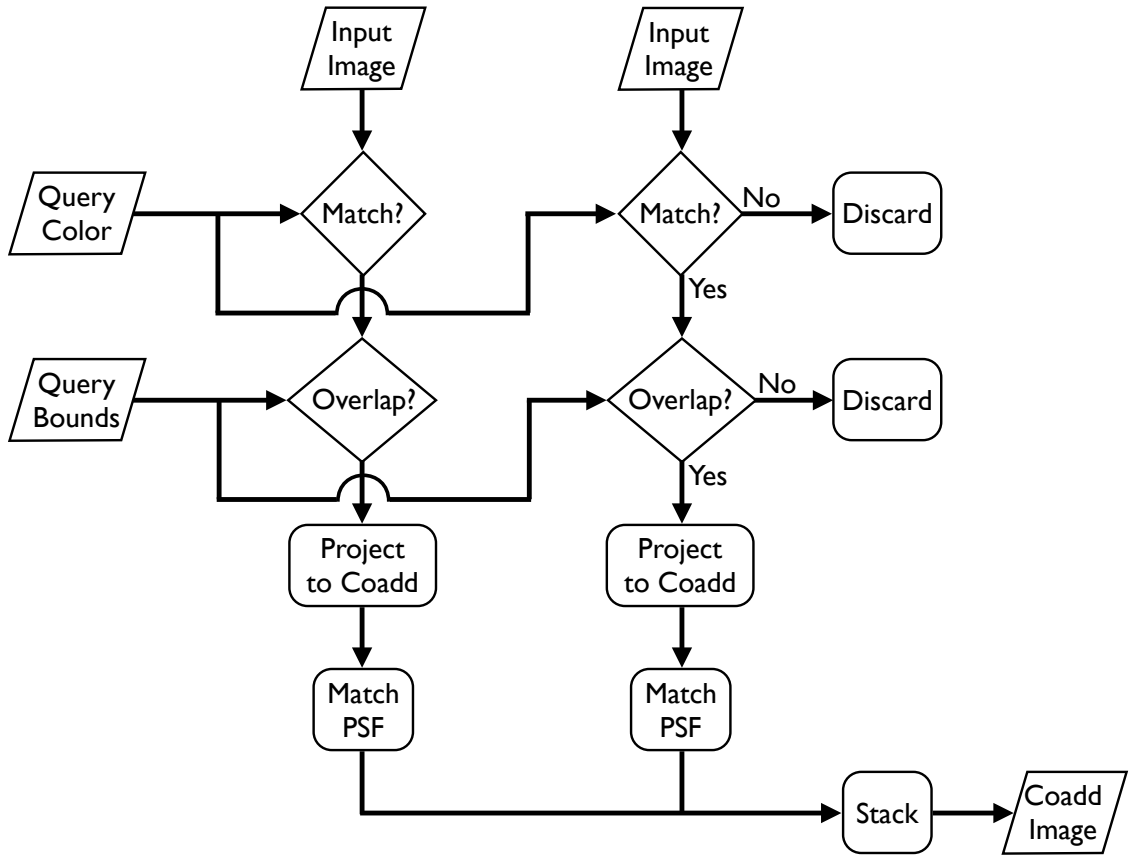}
\caption{\textbf{Image Coaddition Flowchart} --- This figure shows a flowchart of the overall image coaddition process.  The number of input images is effectively unlimited.  Note that multiple independent simultaneous queries are also supported.\label{ImageCoaddition}}
\end{figure}

\begin{figure}
\epsscale{.8}
\plotone{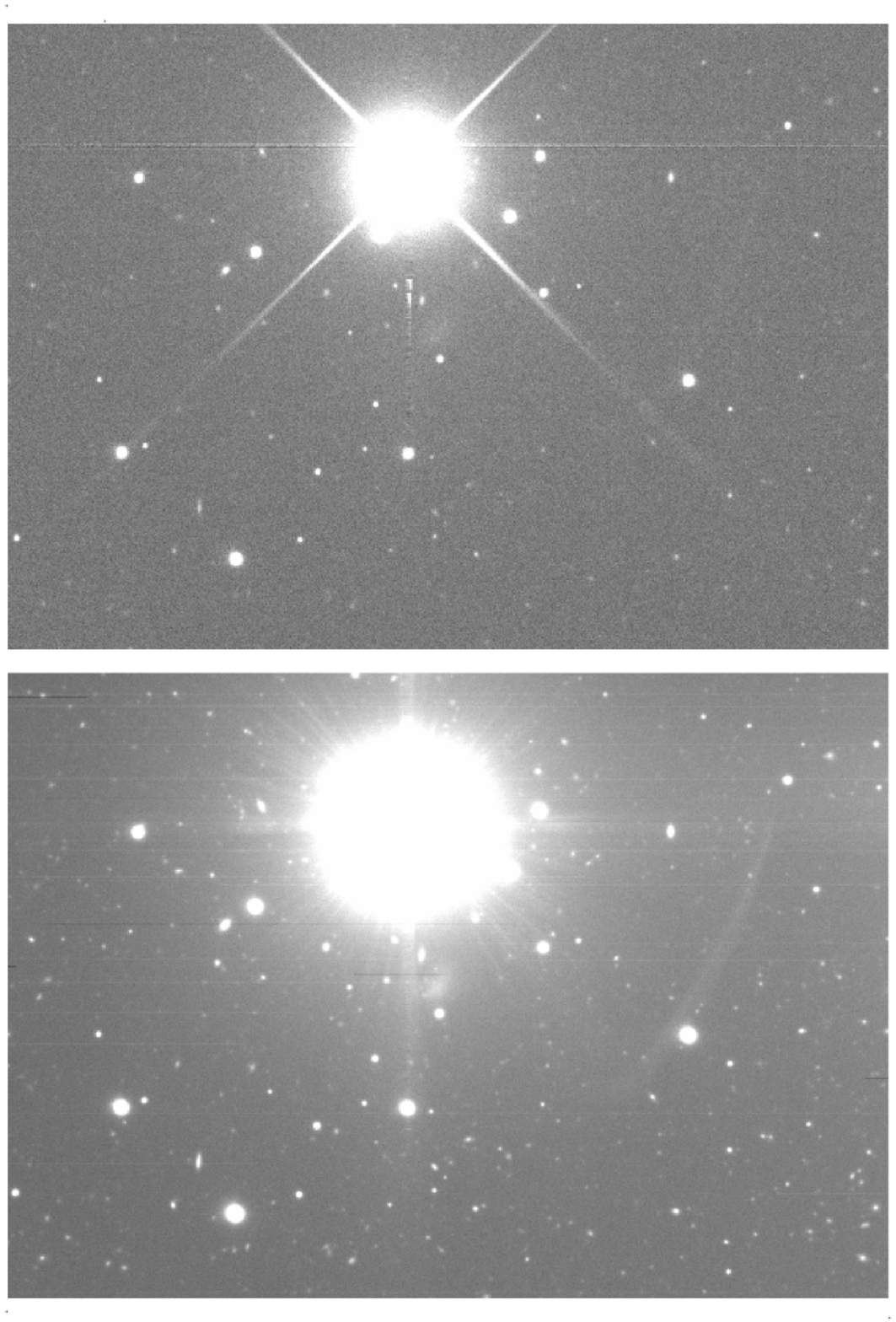}
\caption{\textbf{Image Coaddition Example} --- This figure demonstrates the effect of image-stacking.  Both images show the same region of sky (SDSS Stripe 82, R bandpass) but the image on bottom is a stack of seventy-nine independent exposures.  The SNR is improved by approximately a factor of nine, the observable result of which is a dramatic decrease in the noise (grain), thus revealing numerous sources which are undetectable in the single exposure shown on top.\label{StackingExample}}
\end{figure}

\textit{Image coaddition} (often known simply as \textit{stacking}) is a method of image-processing wherein multiple overlapping images are combined into a single image (see Figs. \ref{ImageCoaddition} and \ref{StackingExample}).   The process is shown in Algorithm \ref{algCoadd}. First, a set of relevant images is selected from a larger database.  To be selected, an image must be in a specified bandpass filter (line 6) and must overlap a predefined bounding box (line 9).  The images are projected to a common coordinate system (line 10), adjusted to a common point spread function (PSF) (line 10), and finally stacked to produce a single consistent image (line 18).

\begin{algorithm}
\begin{algorithmic}[1]
\REQUIRE \{$ims$: one set of images, $qr$: one query\}
\ENSURE \{$coadd$: a single stacked, mosaiced image, $depth$: $coadd's$ per-pixel coverage map\}
\STATE $tprojs \leftarrow$ empty\COMMENT{set of projected intersections}
\STATE $qf \leftarrow$ $qr$ filter\COMMENT{bandpass filter}
\STATE $qb \leftarrow$ $qr$ bounds
\FOR{each image $im$ $\in$ $ims$}
	\STATE $imf \leftarrow$ $im$ filter
	\IF{$imf$ = $qf$}
		\STATE $imb \leftarrow$ $im$ bounds
		\STATE $twc \leftarrow$ intersection of $qb$ and $imb$\COMMENT{Get intersection in RA/Dec}
		\IF{$twc$ is not empty}
			\STATE $tproj \leftarrow$ projection of $im$ to $twc$\COMMENT{Astrometry/interpolation/PSF-matching}
			\STATE Add $tproj$ to $tprojs$
		\ENDIF
	\ENDIF
\ENDFOR
\STATE $coadd \leftarrow$ initialized image from $qr$
\STATE $depth \leftarrow$ initialized depth-map\COMMENT{same dimensions as $coadd$}
\FOR{each image $tproj$ $\in$ $tprojs$}
	\STATE Accumulate $tproj$ illumination into $coadd$
	\STATE Accumulate $tproj$ coverage into $depth$
\ENDFOR
\end{algorithmic}
\caption{Coadd}
\label{algCoadd}
\end{algorithm}

\subsection{Experimental Dataset}
\label{subsecExperimentalDataset}

Much of modern astronomy is conducted in the form of large sky surveys in which a specific telescope and camera pair are used to produce a consistent\footnote{A set of images with similar spatial, color, and noise statistics.} database of images.  One recent and ongoing example is the \textit{Sloan Digital Sky Survey} (SDSS) which has thus far produced seventy terabytes of data \citep{SDSSDR7}.  The next generation survey is the \textit{Large Synoptic Survey Telescope} (LSST) which will produce sixty petabytes of data over its ten year lifetime starting this decade.  These datasets are large enough that they cannot be processed on desktop computers.  Rather, large clusters of computers are required.

\begin{figure}
\includegraphics[scale=.30]{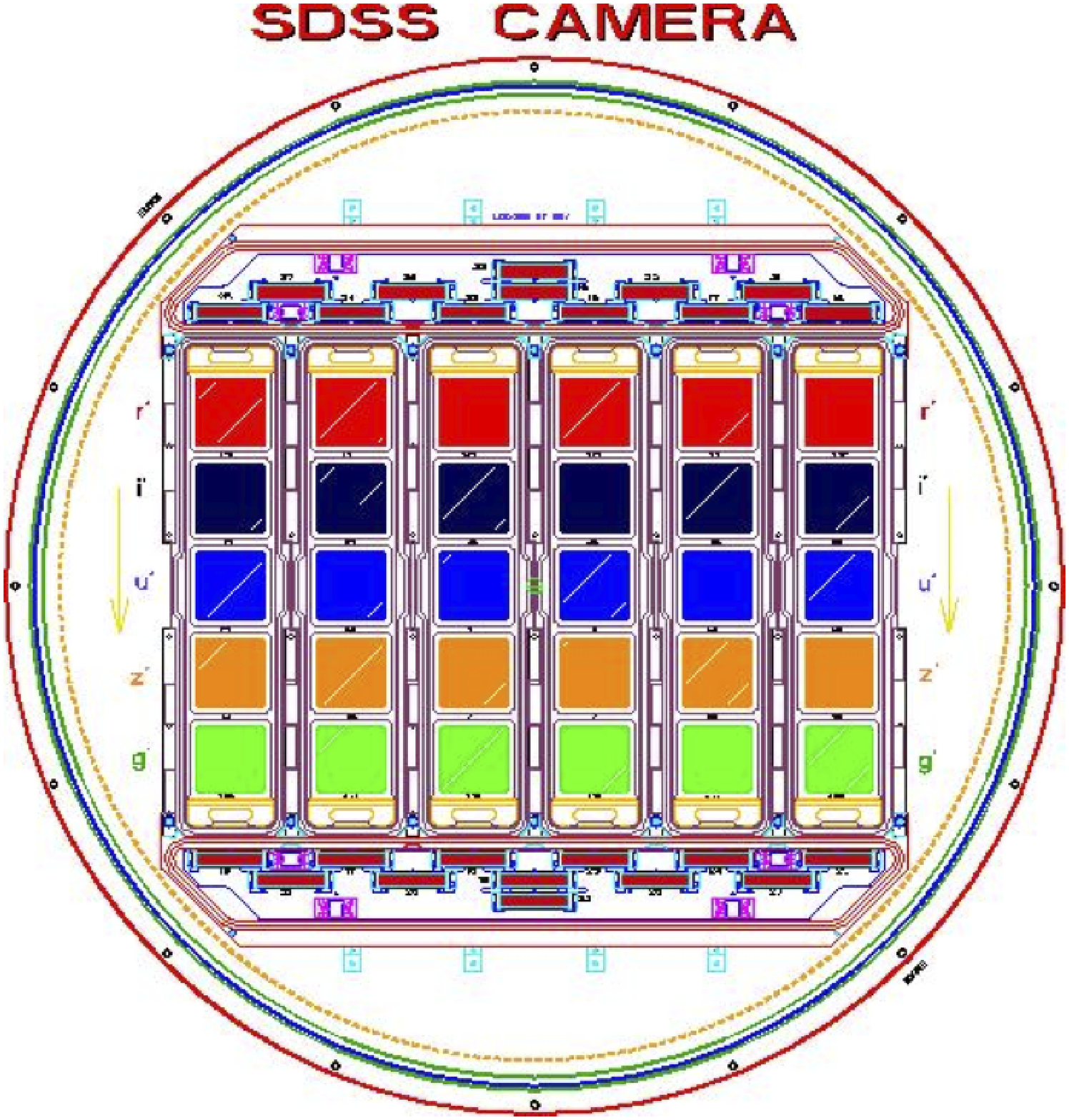}
\caption{\textbf{Sloan Camera CCD Layout} --- The camera used by the SDSS contains 30 CCDs arrange in a grid.  Each row of six CCDs captures light with the same bandpass filter while each column of five CCDs captures a strip of sky that abuts the adjacent strips.\label{SloanCamera}}
\end{figure}

In this paper we consider the SDSS database.  The SDSS camera has 30 CCDs arranged in a 5x6 grid representing five bandpass filters and six abutting strips of sky (see Fig. \ref{SloanCamera}).  Images are captured by drift-scan such that the camera remains fixed during an integration session and the sky moves across the field of view.  The CCD read-out rate is slowed down dramatically (as compared to a typical rapid-capture CCD) so as to shift pixel charge from one row of the CCD to the next in precisely side-real time.  In this way, images are captured without trails yet without moving the telescope.  The resulting integration time per point on the sky is consequently determined by the arc subtended by a CCD's length along the right ascension axis relative to the sky (54.1s).

\subsection{Cluster Configuration and Experiment Dataset}
\label{subsecClusterConfigurationExperimentDataset}

We ran our experiments on a Hadoop-configured cluster.  This cluster is funded by the NSF \textit{Cluster Exploratory} grant (CluE) and is maintained by Google and IBM.  At the time of this research, the cluster consisted of approximately 400 quad-core nodes (more nodes were added later), each node divided evenly between mapper and reducer slots (described in section \ref{secMapReduce}).  Thus, there were approximately 800 mapper and reducer slots each.  It should be noted that other people working on unrelated projects also have access to this cluster for research purposes and due to the unpredictability of their usage at any given time, the cluster's load could vary dramatically.  This situation introduced wide variances into our timing experiments.

\begin{figure}
\epsscale{.80}
\plotone{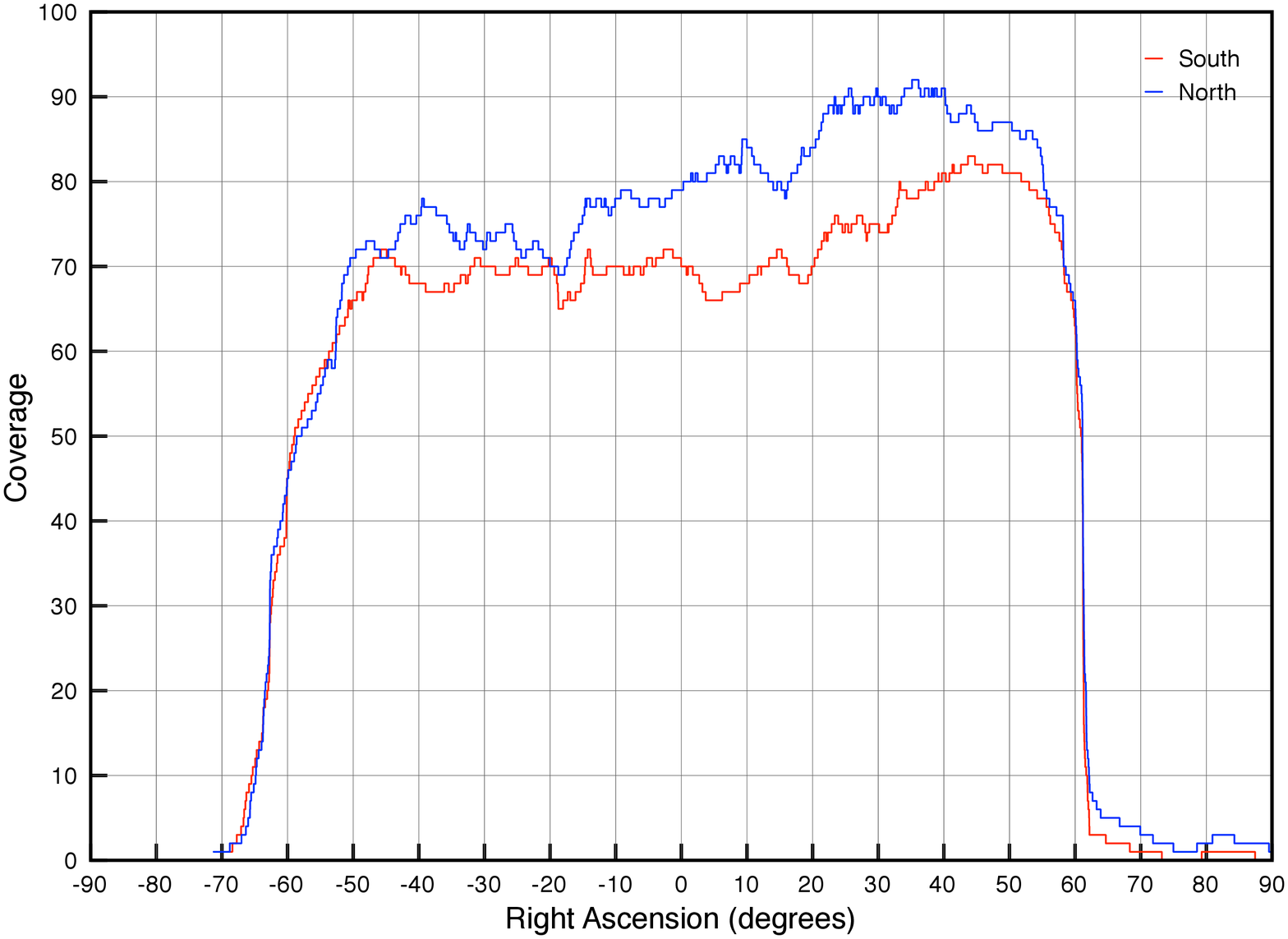}
\caption{\textbf{SDSS Coverage by Right Ascension} --- This plot shows SDSS Stripe 82 coverage as a function of right ascension.  Within its declination range of $+/- 1.25^{\circ}$, average coverage is about 75 between right ascension $-50^{\circ}$ and $+60^{\circ}$.\label{SDSSStripe82RAcoverage}}
\end{figure}

For our experiments we used the Stripe 82 dataset from the SDSS-I (2000-2005)and SDSS-II (2005-2008) surveys.  Since this dataset was captured near the equatorial plane ($+/- 1.25^{\circ}$ declination), the images' world coordinate system mappings are minimally distorted relative to images captured at higher and lower declinations.  Stripe 82 spans right ascension between $-50^{\circ}$ and $+60^{\circ}$ (Stripe 82 only operated in the fall) and has a coverage of about 75 visits (see Fig. \ref{SDSSStripe82RAcoverage}).  Stripe 82 contains about 1,000,000 images in about twenty terabytes.  From this database we defined a three degree long window in right ascension in the most deeply covered region ($37^{\circ}$ to $40^{\circ}$, see Fig. \ref{SDSSStripe82RAcoverage}) and produced the corresponding subset for our experiments, approximately 100,000 images comprising 250GBs of gzip-compressed data (about 600GBs uncompressed).  Using a subset permits us to perform our experiments much more rapidly.  Note that the subset maintains the full depth of the original dataset by being confined to all images within a given spatial bounds.  A more naive method of generating a subset, perhaps by randomly selecting a fraction of the original dataset, would have yielded shallower coadds and any experiments would therefore have offered weaker insight into how the system's performance might scale up to the full dataset.

We ran the Hadoop coadd system described in this paper on two different queries: one representing a lower-bound on the query sky-bounds we expect to be requested when conducting actual research (approximately ${^1/_4}^{\circ}$ square) and the other representing the corresponding upper-bound (approximately $1^{\circ}$ square).  Thus, the two resulting running times demonstrate expected best and worst performance under real-world conditions.

In addition to the two queries that represent our primary experiments, we also briefly considered other queries of identical size but at different locations within the subset's sky-bounds in order to determine the degree to which performance might depend on the specific query.  Most notably the larger query might cover either three or four SDSS strips (camera columns) depending on its precise position along the declination axis.  Such a difference could conceivably have an effect on the performance of the system since it affects the degree to which our prefiltering mechanism (see section \ref{subsubsecPrefiltering}) can eliminate SDSS strips (a filter pass of $^4/_6$ strips in one case and $^3/_6$ strips in the other), thereby imposing a subsequent variation in the number of false positives which must be considered by MapReduce.  Furthermore, due to the general variability of the database, two queries of identical size but located in different locations will almost certainly have some variation in their overall coverage, \ie in the number of FITS images that ultimately contribute to the coadd.  The results of these experiments have suggested that variation in performance resulting from the factors described here is minimal.  The final running time of the overall Hadoop job is virtually unaffected by the choice of query location.  Without loss of generality, we will therefore dispense with any consideration of such variation for the remainder of the paper.

\section{MapReduce}
\label{secMapReduce}

\textit{MapReduce} is a programming model and an associated implementation for parallel data processing \citep{GoogleMapReduce}.  \textit{Hadoop} \citep{ApacheHadoop} is the open-source implementation of MapReduce.  MapReduce has a few primary motivations underlying its design.  First, the overarching goal is to achieve \textit{data locality}, \ie to move the computation to the data rather than move the data to the computation.  Specifically, Hadoop attempts to dispatch each task in the distributed computation to the node that holds the data to process..  Another goal of the MapReduce design is to present a simplified framework for describing a data-processing pipeline such that most general applications can be described within that framework.  Finally, MapReduce is designed to scale to clusters with thousands of machines.  At this scale, failures are the norm rather than the exception.  It thus includes machinery to hide compute-node failures from the user program by managing multiple replicas of all data blocks, automatically restarting tasks that fail, and optionally starting multiple redundant tasks to reduce restart overhead when failures occur.

MapReduce consists of two sequential stages in which units of computation called \textit{mappers} and \textit{reducers}  process the initial and an intermediate representation of the input data respectively.  While these two stages occur sequentially (mappers first), data-processing can be highly parallel \textit{within} each of these two stages.  In the \textit{map} stage, a large input dataset is broken into smaller pieces and distributed to numerous independent mapper objects.  Each map processes its input data, in effect performing a transformation of that data to an intermediate representation.  Following the map stage, the full set of intermediate map outputs is further distributed to the \textit{reduce} stage.  This distribution process is basically a massive distributed sort of the mapper outputs, followed by assignment of the outputs to the reducers for subsequent processing.  Following the mapper and intermediate shuffle stages, the reducers perform a second transformation, the output of which is the output of the entire MapReduce process.  MapReduce may be followed by a final merge stage which groups the numerous reducer outputs into a single result, but this last step is generally a simple concatenation and is therefore not considered an integral step in the overall MapReduce process.  While MapReduce can be performed in more complex ways such as running multiple sequential map stages before the final reduce stage (each map stage's outputs feeding the inputs of the next map stage), in this paper we concentrate on the simpler case of a single map stage followed by a single reduce stage.

The implementation of MapReduce used in this paper is Hadoop.  Hadoop is implemented in Java and, in addition to the general MapReduce framework, incorporates the necessary additional components to build a full MapReduce system, \ie a distributed file system (HDFS), a job scheduler, etc.

\section{Implementing Image Coaddition in MapReduce}
\label{secImplementingImageCoadditionMapReduce}

\begin{figure}
\epsscale{1}
\plotone{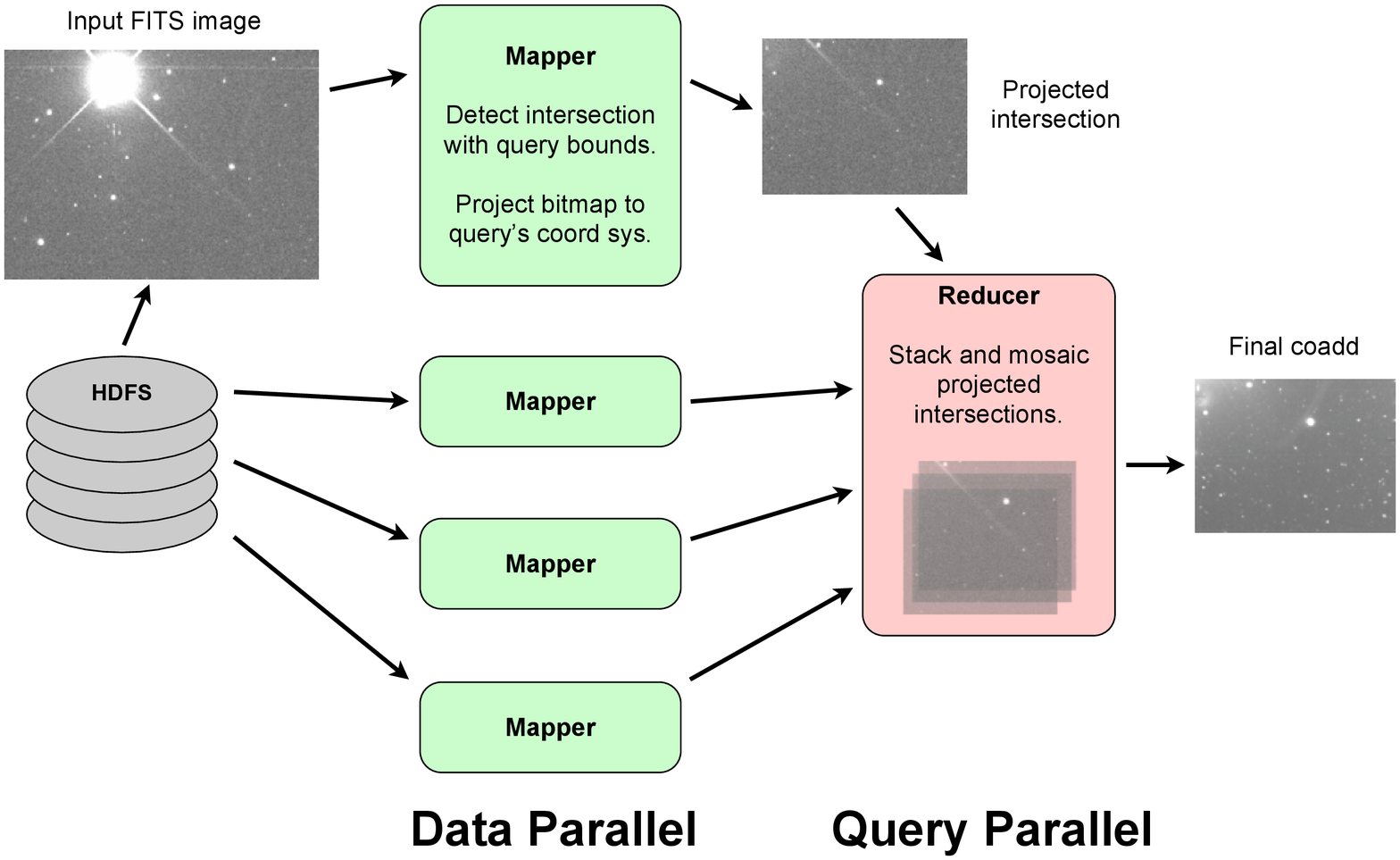}
\caption{\textbf{Image Coaddition in MapReduce} --- This figure shows how we have adapted image coaddition to MapReduce.  Each input image from the database is processed by a single mapper.  Given multiple mappers operating on multiple nodes of the cluster, this stage can be highly parallel.  Each mapper sends a projected bitmap (the intersection of an input image and the query bounds) to a reducer.  The reducer operates in serial for a given query, accumulating all of the projected bitmaps into a final coadded image.  Note however that processing multiple queries is quite feasible.  In such a scenario, the mappers process their input image against multiple queries and send each corresponding projected bitmap to a different reducer.  In this way the reducers operate in parallel over the multiple queries of a given run.\label{imageCoadditionInMapReduce}}
\end{figure}

To adapt a general algorithm to MapReduce we must define two functions, the \textit{mapper} and the \textit{reducer}.  We have adapted image coaddition to MapReduce in the most direct manner possible (see Algorithms \ref{algMap} and \ref{algReduce}).  This straightforward implementation suffices to demonstrate the key benefits for the MapReduce framework as a building block for astronomy image-processing pipelines.  Fig \ref{imageCoadditionInMapReduce} illustrates the overall process.  Each call to the map function processes a single input image.  The image is checked against the query parameters for inclusion in the coadd.  If it passes this test (overlap of the query bounds and designation as the correct bandpass filter) it is then projected to the coadd's coordinate system and the projected bitmap is passed to the reducer.  After all images have been excluded or projected, the reducer then accumulates all of the projected bitmaps into a final coadded image\footnote{Note that in our current work we are not performing PSF-matching, only projection and interpolation.}.

\begin{algorithm}
\begin{algorithmic}[1]
\REQUIRE \{\textbf{Init}: $qr$: one query, \textbf{Key}: unused, \textbf{Value}: $im$: one image\}
\ENSURE \{\textbf{Key}: $qr$, \textbf{Value}: $tproj$: A single image projected to $qr's$ bounds\}
\STATE $imf \leftarrow$ $im$ filter
\STATE $imb \leftarrow$ $im$ bounds
\STATE $qf \leftarrow$ $qr$ filter
\STATE $qb \leftarrow$ $qr$ bounds
\IF{$imf$ = $qf$}
\STATE $twc \leftarrow$ intersection of $qb$ and $imb$\COMMENT{Get intersection in RA/Dec}
	\IF{$twc$ is not empty}
		\STATE $tproj \leftarrow$ projection of $im$ to $twc$\COMMENT{Astrometry/interpolation}
	\ENDIF
\ENDIF
\end{algorithmic}
\caption{Map}
\label{algMap}
\end{algorithm}

\begin{algorithm}
\begin{algorithmic}[1]
\REQUIRE \{\textbf{Key}: $qr$, \textbf{Value}: $tprojs$: A set of images projected to $qr's$ bounds\}
\ENSURE \{\textbf{Key}: unused, \textbf{Value}: unused\}
\STATE $coadd$: A single stacked, mosaiced image
\STATE $depth$: A coverage map corresponding to $coadd$
\STATE $coadd \leftarrow$ initialized image from $qr$
\STATE $depth \leftarrow$ initialized depth-map\COMMENT{same dimensions as $coadd$}
\FOR{each $tproj$ $\in$ $tprojs$}
	\STATE Accumulate $tproj$ illumination into $coadd$
	\STATE Accumulate $tproj$ coverage into $depth$
\ENDFOR
\STATE Write $coadd$ and $depth$ to FITS files\COMMENT{Side-effect of MapReduce process}
\end{algorithmic}
\caption{Reduce}
\label{algReduce}
\end{algorithm}

Algorithm \ref{algReduce} shows that we use a single reducer to perform the summation of the projected intersection bitmaps for a given query, \ie while mappers are parallel over input images, reducers are parallel over queries.  The overall process can readily be extended to process multiple queries at once\footnote{In fact we have implemented the processing of multiple queries, but not thoroughly profiled such behavior.}.  In such cases, the mapper considers the input image against each query and produces a projected bitmap for each query.  The bitmaps are then sent to distinct reducers, each of which performs the summation for a particular query.  Note that the projection and interpolation of the input images into the coadd's coordinate system dominates the computational cost, \ie the summation step of the stacking process in the reducer is a comparatively simple operation.  Furthermore, in our future work we intend to significantly increase the complexity of the mapper operation through the incorporation of more sophisticated coordinate-system projection (warping) and PSF-matching algorithms whereas the reducer's task of pixel summation will remain relatively unaffected.  Therefore, the reducer's serial nature does not hinder the overall running time.  Finally, it should be noted that while the reducer acts in serial on a per query basis, it can act in parallel across queries if a single job processes multiple queries against the overall dataset (see Fig. \ref{imageCoadditionInMapReduce}).  While the output of the reducer is formally a key/value pair our system does not emit the final coadd through this mechanism.  Rather, the coadd is written directly to disk as a side effect of the MapReduce process.

The naive method described in this section summarizes the most straightforward adaptation of image coaddition to MapReduce.  However, it suffers from two inefficiencies.  First, the naive method processes every image once every execution.  Second, finding the location of millions of small individual files during processing dominates the runtime.  In the next section, we address these problems and describe optimizations to solve them.

\subsection{Optimizations}
\label{subsecOptimizations}

\subsubsection{Prefiltering}
\label{subsubsecPrefiltering}

Of the 100,000 FITS files in our experimental dataset, only a small number will ultimately contribute to any one query's coadd.  If we can exclude some of the irrelevant files from consideration before running MapReduce we may improve the overall job's performance.  Therefore, one of the most straightforward ways to improve efficiency is to filter the input and not process the entire dataset.  Filtering can be accomplished by considering the dataset's directory structure and file name convention and subsequently building a partially exclusive \textit{glob} pattern (similar to a regular expression) when specifying the input file set for MapReduce.

\begin{figure}
\includegraphics[scale=.50]{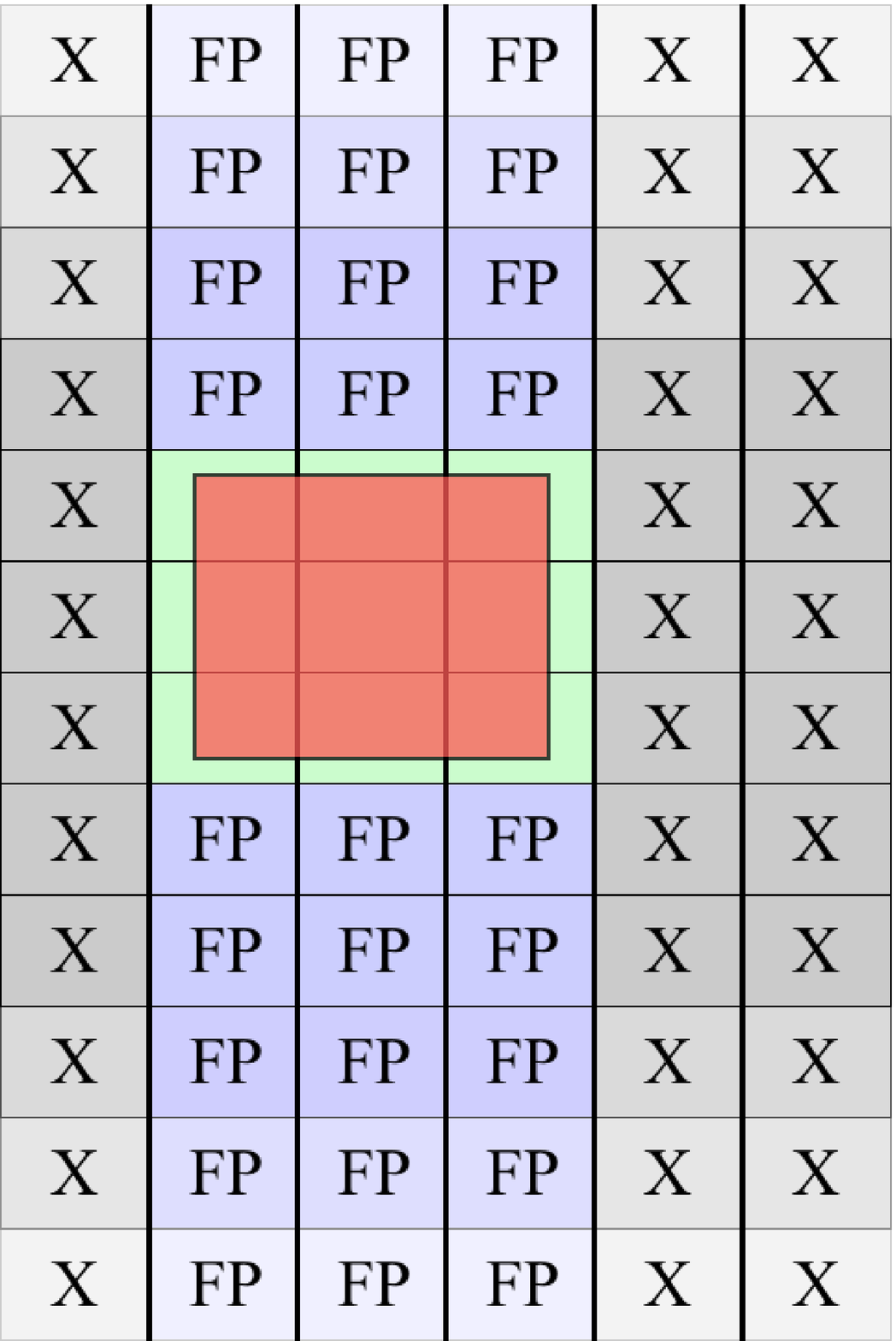}
\caption{\textbf{Single-Axis Spatial Prefiltering} --- In addition to prefiltering images by bandpass filter, we also prefilter images on one spatial axis.  Given that the SDSS camera produces images in six parallel strips of sky (see Fig. \ref{SloanCamera}), we can exclude those strips which do not overlap the query bounds.  This figure shows the spatial layout of a set of FITS files as produced by the camera, arranged into six long strips.  Since the query bounds (inset rectangle) only overlaps columns two, three, and four, we can exclude columns one, five and six (indicated with 'X').  However, the filter suffers from false positives (indicated with 'FP').  Those images will be detected and discarded in the mappers.\label{Prefiltering}}
\end{figure}

Our prefilter works in two ways.  Fig. \ref{SloanCamera} shows a schematic of the SDSS camera which contains an array of 30 independent CCDs.  These 30 CCDs are divided into five rows corresponding to five filters and six columns corresponding to six parallel nonoverlapping strips of the sky.  In the Stripe 82 dataset, these six strips correspond to parallel strips in declination.  Any individual FITS file in the dataset corresponds to an image captured by one of these 30 CCDs.  Since a query explicitly indicates a coadd of a specified bandpass, we can immediately reduce our input dataset by a factor of five by eliminating from consideration those FITS files captured in one of the four other bandpass filters.  Furthermore, the dataset is structured in such a way that is relatively straightforward to filter by CCD column, \ie by spatial location along the declination axis (see Fig. \ref{Prefiltering}).  We perform this secondary filter by determining which of the six columns the query's sky-bounds overlap.  For a very small query we might therefore achieve a further six-fold reduction in the input dataset, although for larger queries this reduction is usually on the order of 1.5 at best (elimination of two of the six total columns).  

The following diagram shows how the SDSS Stripe 82 database is organized.  As an example, let us assume we wish to construct a filter which accepts strips 2, 3, and 4, and bandpass 'g'.  One possible glob\footnote{There are multiple possible ways of constructing a glob which satisfies the filter requirement.  For example, the second '[234]' in the example shown is redundant and could be replaced with a '*'.} would be \textit{/SDSSDB\_ROOT/*/*/corr/[234]/fpC-*-[g][234]-*.fit}.  The diagram below, of the directory hierarchy of the SDSS database, illustrates how this glob pattern can be used to exclude the unnecessary files when building the input set for MapReduce.

\begin{enumerate}
	\item[] SDSSDB\_ROOT
	\begin{itemize}
	\item[] 5902 $\leftarrow$ \textit{SDSS Stripe 82 run id}
		\begin{itemize}
		\item[] 40 $\leftarrow$ \textit{rerun id}
			\begin{itemize}
			\item[] corr  $\leftarrow$ \textit{constant directory name (correlated images)}
				\begin{enumerate}
				\item[] 1 $\leftarrow$ \textit{strip (camera column 1-6)}
					\begin{itemize}
					\item[] fpC-005902-u1-0690.fit $\leftarrow$ \textit{FITS file}
					\item[] fpC-005902-g1-0690.fit
					\item[] fpC-005902-u1-0691.fit
					\item[] fpC-005902-g1-0691.fit
					\end{itemize}
				\item[] 2
					\begin{itemize}
					\item[] fpC-005902-u2-0690.fit
					\item[] fpC-005902-g2-0690.fit
					\item[] fpC-005902-u2-0691.fit
					\item[] fpC-005902-g2-0691.fit
					\end{itemize}
				\item[] 3
				\item[] 4
				\item[] 5
				\item[] 6
				\end{enumerate}
			\end{itemize}
		\end{itemize}
	\end{itemize}
\end{enumerate}

\begin{figure}
\epsscale{.80}
\plotone{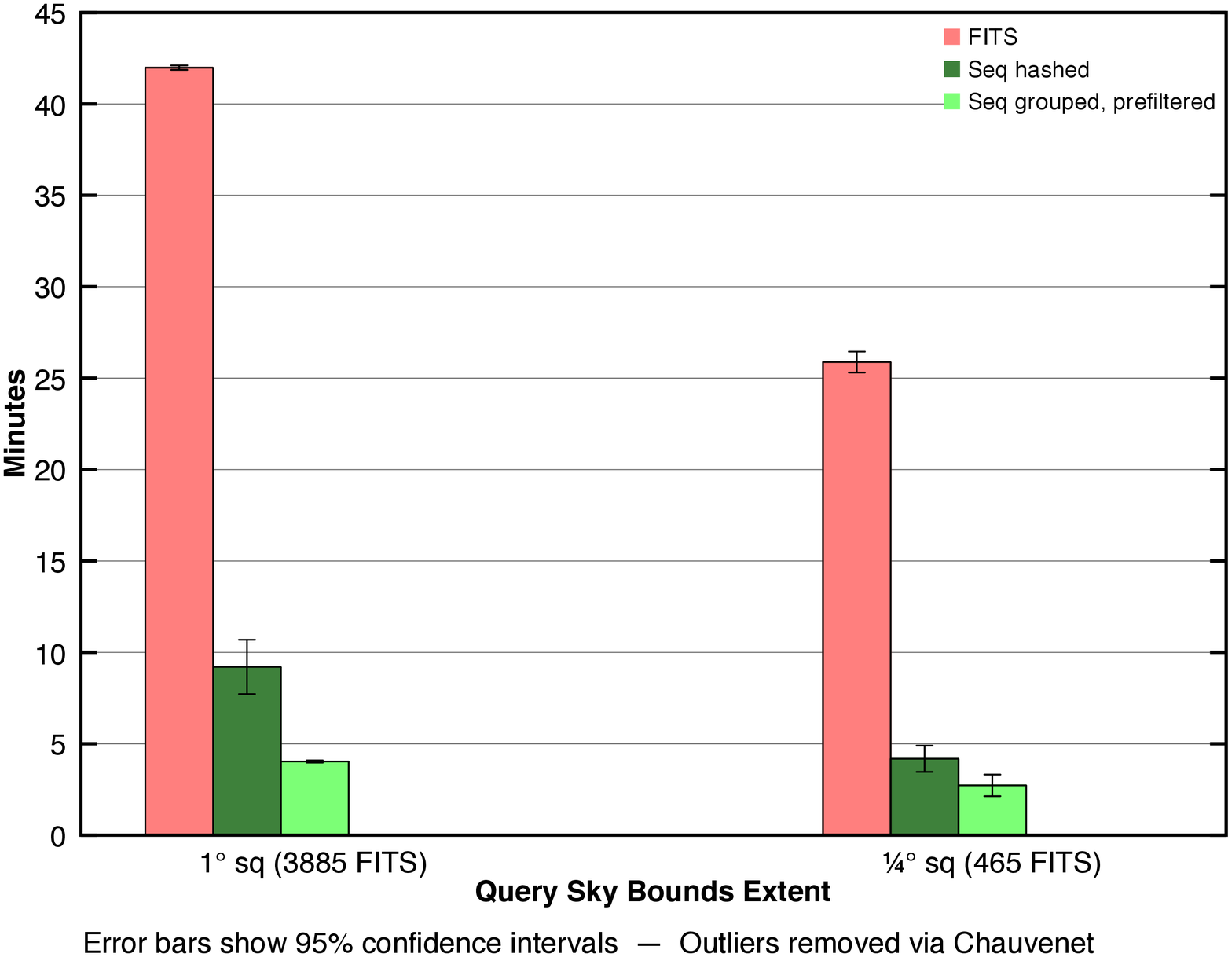}
\caption{\textbf{Coadd Running Times (shorter bars are better)} --- Two queries (large on left, small on right) were run via three methods (from left to right: prefiltered FITS input, non-prefiltered unstructured sequence file input, prefiltered structured sequence file input).  We observe that the second and third methods yielded 5x and 10x speedups relative to the first method for the large query.  Note that the number of mapper input records (FITS files) for the three methods for the large query was 13415, 100058, and 13335 respectively.\label{plot2C}}
\end{figure}

We ran our experiments on two query sizes.  Table \ref{runningTimes} (second row) and Fig. \ref{plot2C} (first bar in each set) show the results of our first round of experiments with the larger query's results in the left column of the table and left plot group.  Note that the spatial filter occurs on only one spatial axis, not two as would be required for a perfect filter.  Given a column that passes the prefilter, all FITS files in the dataset that were captured in that column will be considered by the mappers.  This is unfortunate since many of those FITS files will still not overlap the query bounds in right ascension.  They will be detected via their empty intersections and be discarded by the mappers, but these false positives represent wasted effort.  We discuss this inefficiency and methods for handling it in section \ref{subsubsecUsingSQL}, but first we investigate where the prefiltered method focused its time.

\begin{deluxetable}{lrr}
\tablewidth{0pt}
\tablecaption{Coadd Running Times (minutes) For Two Query Sizes\label{runningTimes}}
\tablehead{
\colhead{Method} &
\colhead{$1^{\circ}$} &
\colhead{${^1/_4}^{\circ}$}
}
\startdata
Raw FITS input, not prefiltered\tablenotemark{a} &		315.0 &	194.0\\
Raw FITS input, prefiltered &						42.0 &	25.9\\
Unstructured sequence file input &					9.2 &		4.2\\
Structured sequence file input, prefiltered &			4.0 &		2.7\\
SQL $\rightarrow$ unstructured sequence file input &	7.8 &		3.5\\
SQL $\rightarrow$ structured sequence file input &		4.1 &		2.2\\
\enddata
\tablenotetext{a}{This method was not explicitly tested.  The times shown were estimated by assuming linearity (see section \ref{subsubsecPrefiltering} for justification) and scaling the second method's running time by the prefilter's reduction factor (7.5), \eg 42.0$\times$7.5 and 25.9$\times$7.5.}
\end{deluxetable}
\clearpage

\begin{figure}
\epsscale{.80}
\plotone{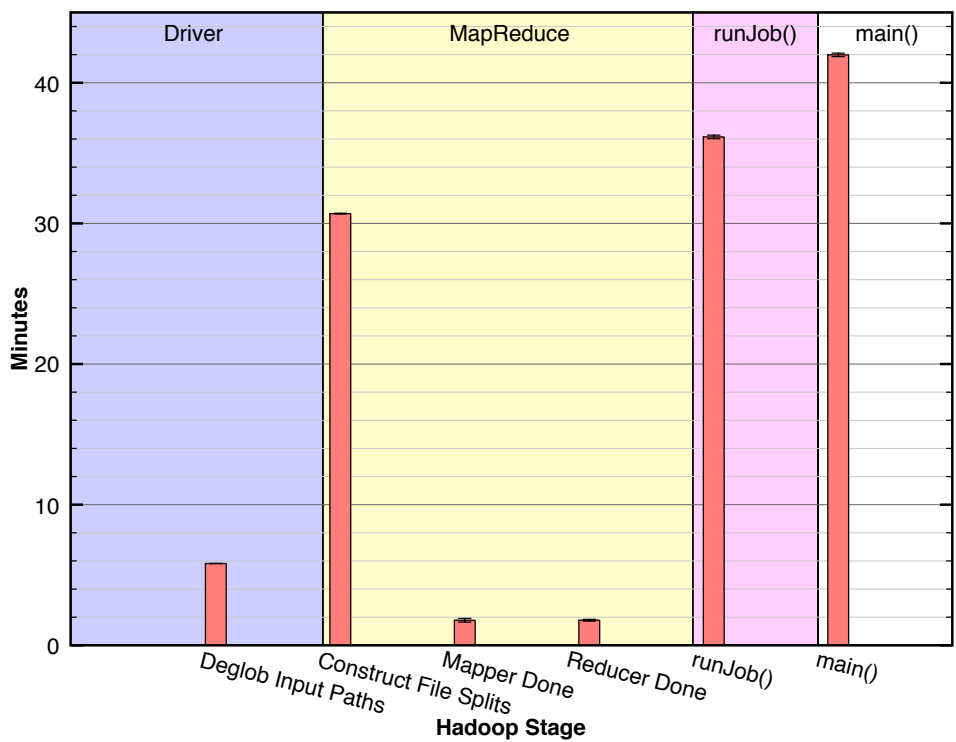}
\caption{\textbf{Running Time Breakdown for FITS Input/Large Query} --- This plot shows a breakdown of the overall running time for the FITS input and large query (see Fig. \ref{plot2C}) into the salient stages.  The last bar represents the total time, 42 minutes.  A call to \textit{main()} involves a preliminary setup stage in the \textit{driver} followed by a single call \textit{runJob()} which encapsulates the MapReduce process.  Thus, the bar in the fourth region is the sum of the bars in the first and third regions.  Likewise, the call to \textit{runJob()} runs the \textit{MapReduce} process and has been further broken down into the components shown in the second region.  Thus, the bar in the third region is the sum of the bars in the second region.  We observe that the dominating step of the overall process is \textit{Construct File Splits}.  This indicates that the Hadoop coadd process spent most of its time locating the FITS files on HDFS and preparing them as input to the mappers.  The actual image-processing computations are represented only in the two shortest bars labeled \textit{Mapper Done} and \textit{Reducer Done} indicating that Hadoop spent a small proportion of its total time on the the fundamental task.  This observation suggests that there is substantial inefficiency in the associated method.\label{plot3FITSonly}}
\end{figure}

Fig. \ref{plot3FITSonly} shows a breakdown of the prefiltered job's overall running time for the larger query into the salient stages.  The last bar represents the total running time, 42 minutes, the same measurement indicated in the first bar of Fig. \ref{plot2C}.  A call to \textit{main()} involves a preliminary setup stage in the \textit{driver} followed by a single call \textit{runJob()} which encapsulates the MapReduce process.  Thus, the bar in the fourth region is the sum of the bars in the first and third regions.  Likewise, the call to \textit{runJob()} runs the \textit{MapReduce} process and has been further broken down into the components shown in the second region.  Thus, the bar in the third region is the sum of the bars in the second region.  We observe that the dominating step of the overall process is \textit{Construct File Splits}.  This indicates that the Hadoop coadd process spent most of its time locating the FITS files on HDFS and preparing them as input to the mappers.  The cost associated with this step results from a set of serial remote procedure calls (RPCs) between the client and the cluster.  The actual image-processing computations are represented only in the two shortest bars labeled \textit{Mapper Done} and \textit{Reducer Done} indicating that Hadoop spent a small proportion of its total time on the the fundamental task.  This observation suggests that there is substantial inefficiency in the associated method, namely in the RPCs involved.  The next section describes one method for alleviating this problem.

We never explicitly measured the performance without prefiltering but an estimate is easy to calculate.  Given that the running times with prefiltering for our two experimental query sizes were 26 and 42 minutes respectively and given that the running time was dominated by the serial RPC bottleneck and given that the prefilter reduced the number of input files by a factor of 7.5 (about 13,000 files out of the 100,000 total passed the prefilter), we can estimate that without prefiltering, the process would have taken approximately 194 and 315 minutes respectively (Table \ref{runningTimes}, first row).  This estimate assumes a linear relationship between the number of files to be found on HDFS and the required time, which is justified given that the number of RPCs is constant per file and the amount of time required per RPC is constant.

\subsubsection{Optimizing Hadoop with Sequence Files}
\label{subsubsecSequenceFiles}

While Hadoop is designed to process very large datasets, it performs better on a dataset of a given size if the total number of files is relatively small.  In other words, given two datasets of identical data with one dataset stored in a large number of small files and the other stored in a small number of large files, Hadoop will generally perform better on the latter dataset.  While there are multiple contributing factors to this variation in behavior, one significant factor is the number of remote procedure calls (RPCs) required in order to initialize the MapReduce job, as illustrated in the previous section.  For each file stored on HDFS, the client machine must perform multiple RPCs to the HDFS namenode to locate the file and notify the MapReduce initialization routines of its location.  This stage of the job is performed in serial and suffers primarily from network latency.  For example, in our experimental database we processed 100,000 files.  Each file had to be independently located on HDFS prior to running the MapReduce job and this location was performed serially over set of 100,000 files.  Consequently, this step alone could take nearly five hours on our setup.  Even in the prefiltered method described above, this step still took about 30 minutes (see Fig. \ref{plot3FITSonly}).

The above is a well-known limitation of the Hadoop system \citep{White} and Hadoop includes an API and a set of data structures specifically targeted at alleviating it.  The relevant structure is a \textit{sequence file}.  A sequence file contains an arbitrary number of key/value pairs where a key is used to indicate (and locate) a specific file and the value is the file in question.  In this way, individual files can be retrieved from a sequence file in a random-access fashion.  In effect, a sequence file is a concatenation of smaller files bundled with an index.

We use sequence files by grouping many individual FITS files into a single sequence file, indexed simply by the FITS filename itself.  Thus, we can retrieve a specific FITS file rapidly while simultaneously reducing the number of actual files on disk from the number of FITS files to the number of resulting sequence files.  As a first strategy, we randomly grouped FITS files into sequence files (thus producing \textit{unstructured} sequence files), as illustrated in Fig. \ref{sequenceFiles} (top).  Table \ref{runningTimes} and Fig. \ref{plot2C} show the results of using this approach in row three of the table and bar two of each plot group.  As the results show, this approach improved upon the prefiltered method (described above) by a factor of five and theoretically improved upon the most naive approach by a factor of approximately 38.  It should be noted that we never implemented the most naive method and so have no running times for it.  Consequently, the prefiltered method as applied directly to the FITS database should be taken as the baseline for performance comparisons.

Note that unstructured sequence files outperform prefiltering of the direct FITS file dataset by a factor of five.  Table \ref{runningTimes} and Fig. \ref{plot2C} illustrate this fact.  What is interesting is that this five-fold improvement occurs \textit{despite} the fact that in the prefiltered case we actually reduced the input of 100,000 FITS files down to about 13,000 files while in the sequence file case we could not perform any prefiltering.  In the unstructured sequence file case, many more FITS files were processed relative to the prefiltered FITS file case.  Despite this inefficiency on part of the sequence file method, it still significantly outperforms the prefiltering method due to its vast reduction in the number of actual files on disk, \ie the conversion of the input representation from a set of FITS files to a set of sequence files.

\subsubsection{Unstructured and Structured Sequence Files}
\label{subsubsecUnstructuredSequenceFiles}

FITS files can be assigned to sequence files in any way the programmer chooses.  In the previous section, we considered the worst case performance by using an unstructured sequence file database.  The resulting sequence file database could not be pruned in any meaningful fashion at runtime, \eg the prefiltering mechanism described in section \ref{subsubsecPrefiltering}, and therefore had to be fully read and processed during a MapReduce job.  Results of using such a sequence file database were presented in the previous section.

\begin{figure}
\includegraphics[scale=.48]{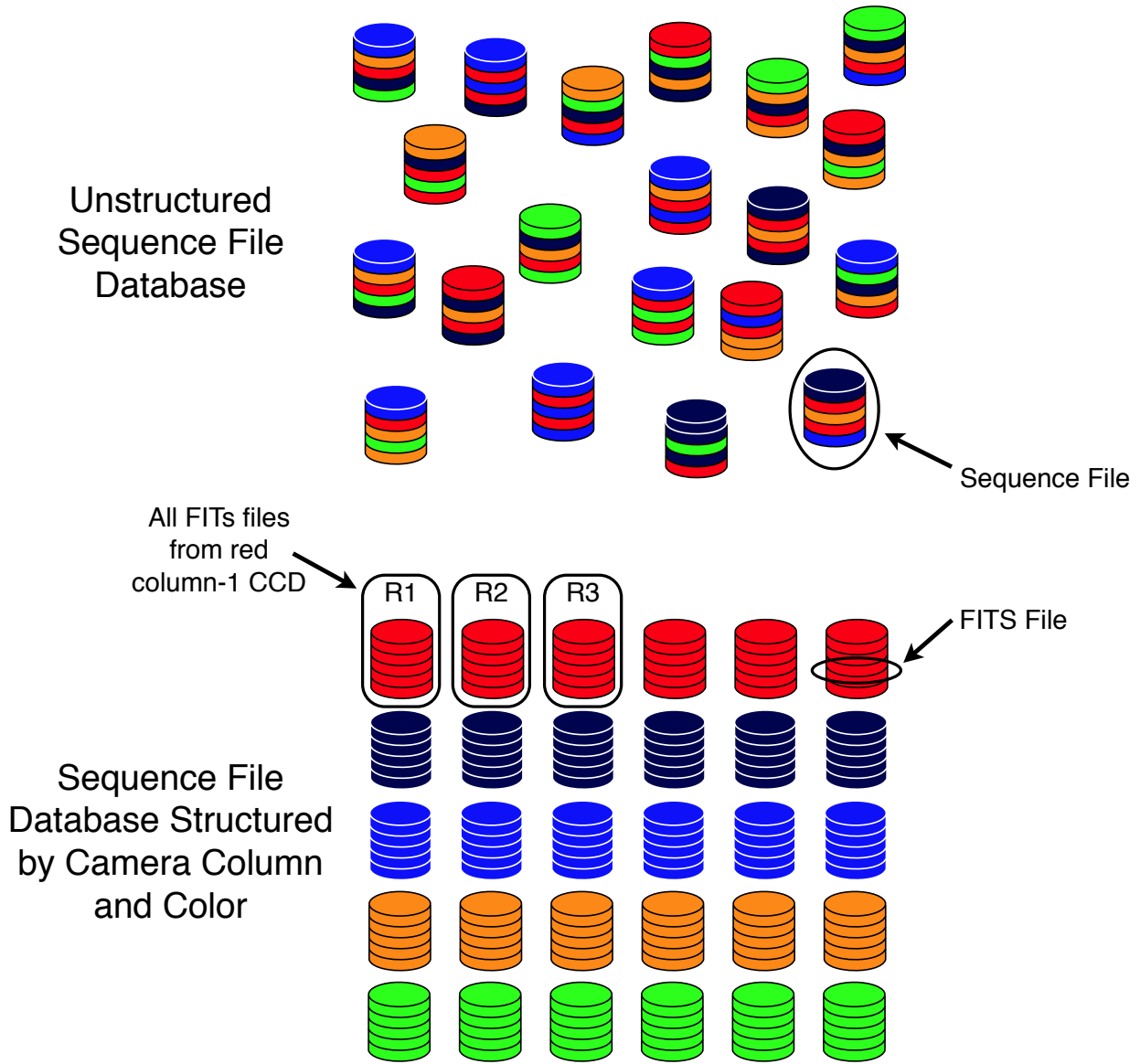}
\caption{\textbf{Sequence Files} --- A sequence file (each column) comprises a set of FITS files (layers).  We convert a database of numerous FITS fits to a database of fewer actual files (sequence files) using one of two methods.  The first method has no structure (top).  FITS files are assigned to sequence files at random.  The second method maps the SDSS camera's CCD layout (see Fig. \ref{SloanCamera}) onto the sequence file database, \ie there is one sequence file for each CCD on the camera.  FITS files are assigned accordingly and thus a given sequence file contains FITS files of only one bandpass and from only one column of the Camera.\label{sequenceFiles}}
\end{figure}

However, it is also worth considering a more rationally motivated sequence file structure since doing so might permit pruning, \ie filtering as demonstrated above.  We refer once again to Fig. \ref{SloanCamera} which shows the SDSS camera's CCD layout.  This time, we use the SDSS camera's CCD arrangement not to define a prefiltering method on FITS files, but rather to impose a corresponding structure on the sequence file database.  We define 30 distinct sequence file types, one corresponding to each CCD of the camera (see Figs. \ref{SloanCamera} and \ref{sequenceFiles}).  Thus, a given sequence file contains only FITS files originating from one of the original 30 CCDs.  If the sequence files are named in a way which reflects the glob filter described previously, we can then filter entire sequence files in the same way that we previously prefiltered FITS files.  This method of filtering permits us to eliminate entire sequence files from consideration prior to MapReduce on the basis of bandpass or column coverage of the query.  We anticipated a corresponding improvement in performance resulting from the reduction of wasted effort spent considering irrelevant FITS files in the mappers.

Table \ref{runningTimes} (row four) and Fig. \ref{plot2C} (third bar in each set) show the results of using structured sequence files and prefiltering in concert.  We observe a further reduction in running time over unstructured sequence files by a factor of two on the larger query.  The savings in computation relative to unstructured sequence files resulted from the elimination of much wasted effort in the mapper stage considering and discarding FITS files either whose bandpass did not match the query or whose sky-bounds did not overlap the query bounds.

Both prefiltering methods suffer from false positives resulting from the fact that the spatial filter occurs on only one spatial axis.  Likewise, the non-prefiltered method performs no prefiltering at all.  Therefore, in all three methods, effort is still being wasted considering irrelevant FITS files.  The prefiltering methods processed 13,000 FITS files in the mappers while the non-prefiltered method processed the entire dataset of 100,000 FITS files.  However, the number of FITS files that actually contributed to the final coadd was only 3885.  In the next section we describe how we used a SQL database and query prior to running MapReduce to eliminate this problem.

\subsubsection{Using a SQL Database to Prefilter the FITS Files}
\label{subsubsecUsingSQL}

\begin{figure}
\includegraphics[scale=.28]{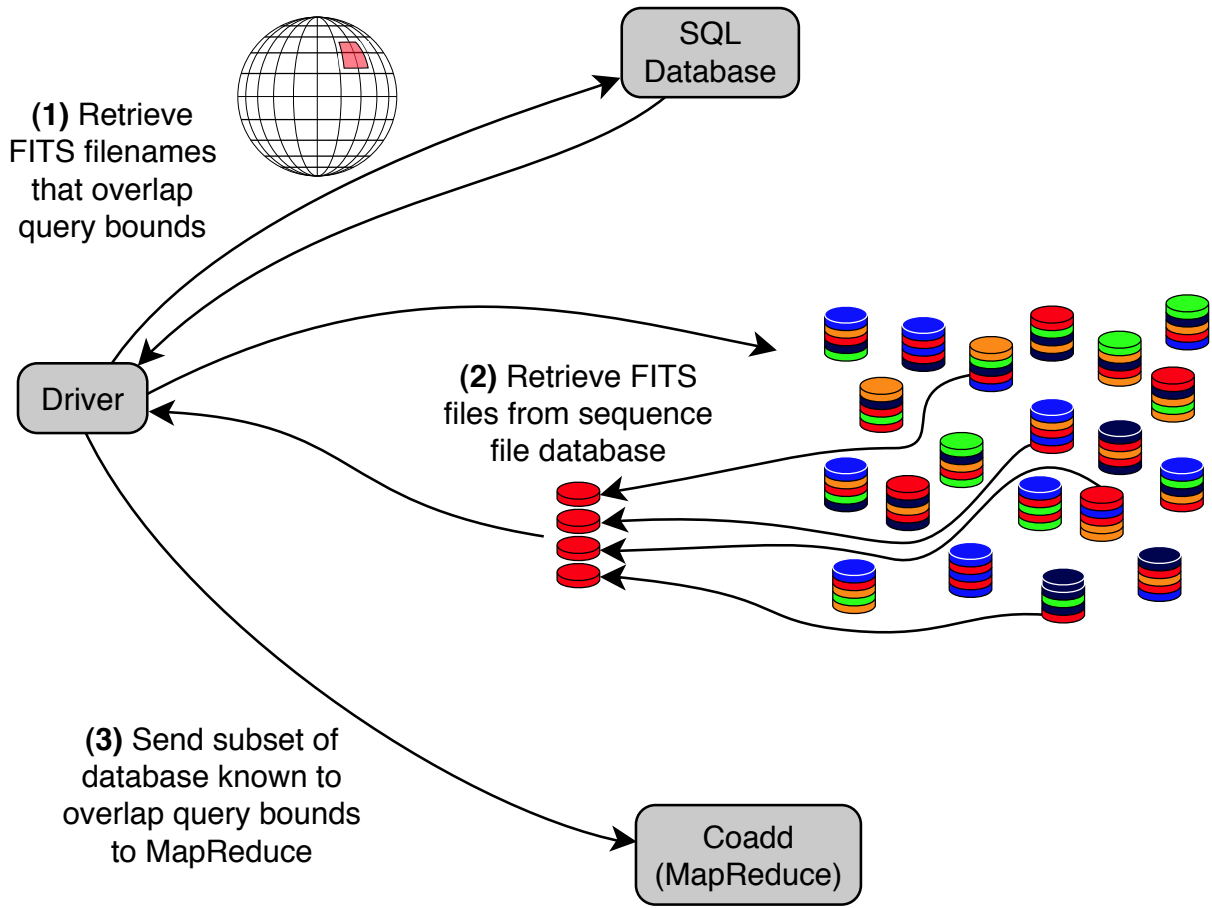}
\caption{\textbf{SQL Utilization} --- SQL is used as method of prefiltering prior to running MapReduce.  The user's query (bandpass and sky-bounds) is used to generate an SQL query to retrieve the names of the relevant FITS files \textbf{(1)}.  Those FITS files are then retrieved from the sequence file database \textbf{(2)} and finally are sent to MapReduce \textbf{(3)}.  In this way, MapReduce only receives FITS files as input which are guaranteed to be relevant to the image coaddition process.  Clearly, either of the two sequence file databases described can be used (see Fig. \ref{sequenceFiles}).  Please see the text for a comparison of how the performance is affected by this option.\label{SQLCoadd}}
\end{figure}

All three methods described previously (prefiltered FITS files, unstructured sequence files, and prefiltered structured sequence files) process irrelevant FITS files in the mappers.  Toward the goal of alleviating this inefficiency, we devised a new method of prefiltering.  This method consists of building a SQL database of the FITS files outside of HDFS and outside of Hadoop in general.  The actual image data is not stored in the SQL database --- only the bandpass filter and sky-bounds of each FITS file are stored, along with the necessary HDFS file reference data necessary to locate the FITS file within the sequence file database, \ie its assigned sequence file and its offset within the sequence file.  Running a job using this method consists of first performing a SQL query to retrieve the subset of FITS files that are relevant to the user's query, and from the SQL result constructing a set of HDFS \textit{file splits}\footnote{Although the concept of a \textit{file split} is complicated when one considers the full ramifications of storing files on a distributed file system, it suffices for our discussion to define a file split simply as the necessary metadata to locate and retrieve a FITS file from within a host sequence file.} to specify the associated FITS files within the sequence file database (See Fig. \ref{SQLCoadd}).  The full set of file splits then comprises the input to MapReduce.

The consequence of using SQL to determine the relevant FITS files and sending only those FITS files to MapReduce is that this method does not suffer from false positives as described above.  Thus, the mappers waste no time considering (and discarding) irrevelant FITS files since every FITS file received by a mapper is guaranteed to contribute to the final coadd.  The intention is clearly to reduce the mapper running time as a result.  Table \ref{numFitsProcessed} shows the number of FITS files read as input to MapReduce for each of the six experimental methods.  Note that prefiltering is imperfect, \ie it suffers from false positives and accepts FITS files which are ultimately irrelevant to the coaddition task.  However, the SQL methods only process the relevant files.

\begin{deluxetable}{lrr}
\tablewidth{0pt}
\tablecaption{Number of FITS Files Processed by MapReduce\label{numFitsProcessed}}
\tablehead{
\colhead{Method} &
\colhead{$1^{\circ}$}(3885\tablenotemark{a}) &
\colhead{${^1/_4}^{\circ}$}(465\tablenotemark{a})
}
\startdata
Raw FITS input, not prefiltered\tablenotemark{b} &		100058 &	100058\\
Raw FITS input, prefiltered &						13415 &	6714\\
Unstructured sequence file input &					100058 &	100058\\
Structured sequence file input, prefiltered &			13335 &	6674\\
SQL $\rightarrow$ unstructured sequence file input &	3885 &	465\\
SQL $\rightarrow$ structured sequence file input &		3885 &	465\\
\enddata
\tablenotetext{a}{Indicates coverage for each query (the number of FITS files from the dataset that overlap the query bounds).}
\tablenotetext{b}{This method was not explicitly tested, but since it performs no prefiltering, it should be clear that the entire dataset would be processed.}
\end{deluxetable}
\clearpage

\begin{figure}
\epsscale{.80}
\plotone{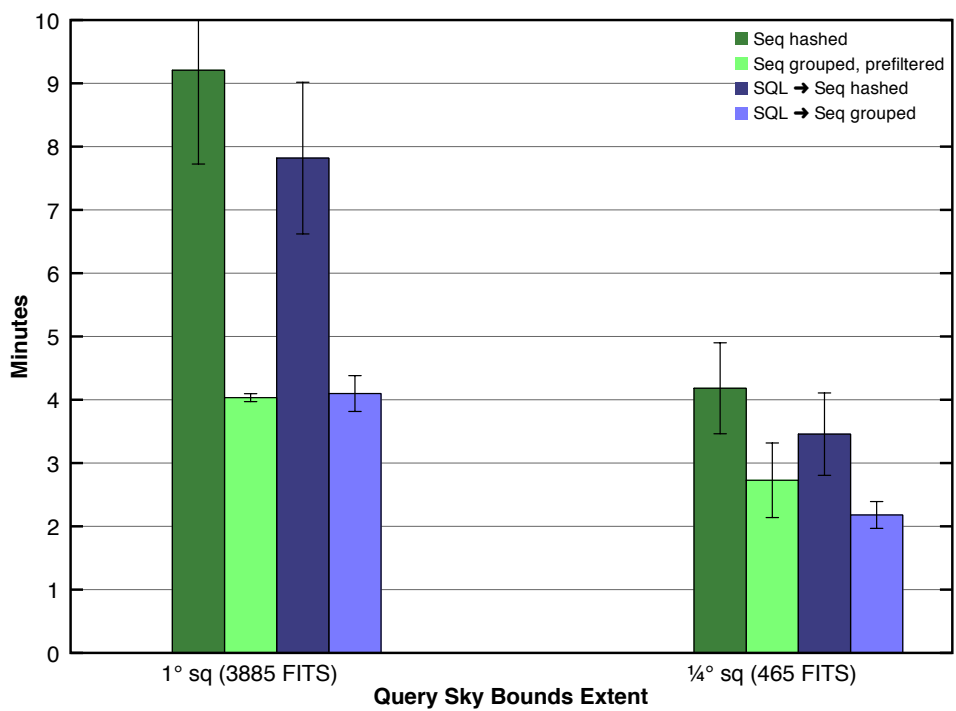}
\caption{\textbf{Coadd Running Times (shorter bars are better)} --- Two queries (large on left, small on right) were run via four methods (from left to right: non-prefiltered unstructured sequence file input, prefiltered structured sequence file input, SQL unstructured sequence file input, SQL structured sequence file input).  The first two bars in each set correspond to the last two bars from Fig. \ref{plot2C} while the second two bars provide comparative running times using the SQL method on each of the two sequence file types.  For any given sequence file type (bars one and three or bars two and four) we observe little improvement resulting from the SQL method despite the inefficiency suffered by the nonSQL method (consideration and discarding of irrelevant FITS files in the mappers).  Furthermore, we observe that SQL performs quite differently on the two sequence file types (compare bars three and four) despite the fact that the two methods process an identical set of FITS files and perform an identical set of computations.  Please see the text for an explanation of this variation in performance.\label{plot4}}
\end{figure}

Fig. \ref{plot4} shows the results of the new method.  We have removed the prefiltered FITS method from the plot (shown in Fig. \ref{plot2C}) so that we may concentrate our attention on the faster methods, namely, the two sequence file methods previously described (the first two bars shown in Fig. \ref{plot4}), and two new methods, one each for performing SQL against the unstructured sequence file database (the third bar) and for performing SQL against the structured sequence file database (the fourth bar).

Several patterns are observed in the plot.  First, note that it is only meaningful to compare SQL vs. nonSQL for a given sequence file database, either structured or unstructured.  Thus, the appropriate comparisons to make are between bars one and three or between bars two and four.  We observe that in such comparisons, the SQL method does successfully outperform the nonSQL method in most cases, but to a lower degree than we hoped for when we implemented it, \ie bar three shows only a small improvement over bar one and bar four shows virtually no improvement over bar two.

Let us consider the results indicated by bars two and four for the larger query.  The noteworthy observation is that while SQL shows some benefit, that benefit is lower than originally anticipated given the 3.5x difference in mapper input records (FITS files), 13335 vs. 3885.  13335 is the number of FITS files read in by the prefiltered sequence file method (of which 3885 were relevant and the rest discarded) while 3885 is the number of FITS files read in by the SQL method (all of which were relevant).  Therefore, we can conclude that in this case the cost of considering and discarding numerous irrelevant FITS files was negligible and likewise that the additional complexity imposed by supporting and using an external SQL database offers no benefit.

Let us consider the results indicated by bars one and three for the larger query.  Again, we note little improvement in performance given the variance of the confidence intervals.  This is surprising given the dramatic difference in mapper input records, a full 26 fold difference.  The first bar represents non-prefiltered unstructured sequence files, thus 100,058 input records while the third bar represents a SQL method, thus 3885 input records.  The conclusion is similar in this case, that the cost of discarding irrelevant files is low.

We might then ask: why is there a two-fold difference in performance between the two SQL methods considering that they both processed exactly the same amount of data?  To answer this question we must investigate not merely the number of mapper input records but the number of \textit{mapper objects} launched by Hadoop.  These two values are rarely equal because a mapper can be reused to process multiple input records.  In the SQL-unstructured-sequence-file method, the 3885 input FITS files were processed by 1714 mapper objects (about two FITS files each) while in the SQL-structured-sequence-file method, 338 mapper objects were used (about eleven FITS files each). This discrepancy is due to the way we assign FITS files to each mapper object. For each mapper object, we assign FITS files in the same HDFS block.  Due to the replication of blocks across HDFS, copies of each block will be stored on multiple hosts. When possible, Hadoop will schedule mapper objects to run on one of the hosts where that mapper's input blocks are located so that the files can be efficiently accessed from local disks.  Clearly, this approach generates a variable number of mapper objects depending on the structure of the input sequence file.  In the case of unstructured sequence files, the FITS files relevant to a given query are scattered throughout the sequence file database whereas in the case of structured sequence files, the relevant FITS files are more tightly packed.  Thus, greater data locality is achieved in the assignment of FITS files to mappers in the structured case and fewer mapper objects are required.  To complete our understanding of the difference in running times we must further consider the maximum number of mapper objects that the cluster can sustain simultaneously.  Since a mapper represents a conceptual unit of computation, it corresponds to --- or directly relies upon --- a processor core in order to operate.  Therefore, the number of simultaneously sustainable mapper objects for a Hadoop job is limited by the number of cores available on the entire cluster \footnote{Ignoring for the moment additional factors such as distribution of core assignments between mapper and reducer objects and shared usage of the cluster by concurrent MapReduce jobs unrelated to our application.}.  In our case, this value is about 800.  We can now see one fundamental problem with the unstructured case: not all of the 1714 mappers could run simultaneously; some could not begin processing until after others had completed their own processing.  This limitation was not true in the case of structured sequence files where only 338 slots were required and they could all run simultaneously.  However, one might actually predict superior performance from the unstructured case for the simple fact that it benefits from 800x parallelism while the structured case only benefits from 338x parallelism.  We theorize that the explanation for why such behavior was not observed lies in the nonnegligible startup cost of launching a mapper object, \ie there is a genuine benefit in reusing mappers and there must be some tipping point in this tradeoff where the benefit of additional parallelism is outweighed by the cost of creating mappers for brief computational needs.

\section{Conclusions}
\label{secConclusions}

This work presented our implementation and evaluation of image coaddition within the MapReduce data-processing framework using Hadoop.  We investigated five possible methods of implementation with the latter methods designed to improve upon the earlier methods.

Our first round of experiments processed a dataset containing 100,000 individual FITS files.  Despite the use of a prefiltering mechanism to decrease the input size, this method yielded poor performance due to the job initialization costs.  To decrease the initialization time we must process a dataset consisting of fewer actual files, but obviously without altering the underlying data.  This goal is achieved through the use of sequence files which group individual files together.  Sequence files are offered through the Hadoop API for precisely this purpose.

Our next round of experiments considered sequence files which were grouped in an unstructured manner and which therefore could not be prefiltered.  Despite this weakness relative to the first method, the use of unstructured sequence files still yielded a five-fold improvement in performance.

We then ran experiments on sequence files which were grouped in ways that reflected the same prefiltering mechanism used earlier.  The sequence files themselves could then be prefiltered.  This method yielded another two-fold improvement in performance (ten-fold over the original method).

The first three methods described all suffered from an inefficiency whereby irrelevant FITS files were considered and discarded in the mapper stage.  Toward the goal of alleviating this inefficiency we devised our fourth and fifth methods, which used a SQL query prior to MapReduce to identify the relevant FITS files.  Those FITS files were then retrieved from the two previously described sequence file databases to run MapReduce.  When applied to the unstructured sequence file database, this SQL method yielded minimal improvement in performance.  The lack of structure of the sequence files prevented the efficient reuse of mapper objects.  However, when the SQL method was used in conjunction with the structured sequence file database, the mappers were reused more efficiently and good performance was achieved, although it should be noted that the SQL method still did not significantly outperform the nonSQL method in which bandpass filter and one-dimensional spatial prefiltering was used.  In the case of the smaller query there was some advantage however.

It should be noted that while SQL showed little improvement over prefiltering in our current experiments, we anticipate that SQL should show significant advantage on larger datasets.  In such a case, prefiltering would be more adversely affected by the false-positive problem, \ie more irrelevant FITS files would pass the prefilter to be considered and discarded by the mappers.  There ought to be a dataset size above which this wasted effort begins to show measurable degradation in performance.  On the other hand, the SQL method should show comparatively more gradual decrease in performance as the dataset grows since it only processes the relevant FITS files to begin with.  This prediction is admittedly speculative as we have not yet conducted the necessary experiment.

The use of Hadoop shows great promise for processing massive astronomical datasets.  On a 400-node cluster we were able to process 100,000 files (300 million pixels) in three minutes using two different methods, one with SQL and one without.  This is a marked improvement over other systems such as Montage, which processes approximately 40 million pixels in 32 minutes \citep{Montage2}.  Even if one considers only the FITS files that were relevant to a query (3885 files, or 12 million pixels), when the three minute running time is scaled by ten to compare to Montage's 32 minutes we get 120 million pixels, still a marked improvement over 40 million.

\section{Future Work}
\label{secFutureWork}

Our current work represents merely the early stages of our research into the use of Hadoop for image coaddition.  Areas of future work include:

\begin{itemize}
\item Conversion of the mapper and reducer code to C++ and incorporation of our existing C++ library of image-processing routines which are capable of performing much more sophisticated coaddition algorthms.
\item The addition of time-bounds to the query parameters so that coadds may be generated only within a specified window of time.  This behavior will enable the investigation of time-variable phenomena such as variable stars and moving objects, \eg asteroids.
\item The addition of subsequent pipeline stages to process the resulting coadds, such as the detection and tracking of moving and variable objects.
\item The development of new parallel machine-learning algorithms for anomaly detection and classification.
\end{itemize}

Ultimately, we intend to develop our system into a full-fledged data-reduction pipeline for petabyte astronomical datasets as will be produced by next-generation sky surveys such as LSST.

\section{Acknowledgements}
\label{secAcknowledgements}

\acknowledgments

This work is funded by the NSF Cluster Exploratory (CluE) grant (IIS-0844580) and NASA grant 08-AISR08-0081.  The CluE cluster is funded through the CluE grand and maintained by IBM and Google.  We thank them for their continued support.  We further wish to thank both the LSST group in the astronomy department and the database research group in the computer science department at the University of Washington.  All members of both groups contributed greatly through feedback, suggestions and draft revisions.

{\it Facilities:} \facility{SDSS}, \facility{LSST}.

\clearpage

\clearpage


\begin{thebibliography}{}
\bibitem[HDF(1999)]{HDF} Szalay, A. S., Connolly, A. J. and Szokoly, G. P.  Simultaneous Multicolor Detection of Faint Galaxies in the Hubble Deep Field.  \textit{The Astronomical Journal}, Jan 1999.
\bibitem[Google(2004)]{GoogleMapReduce} Dean, J. and Ghemawat, S. MapReduce: Simplified Data Processing on Large Clusters.  In \textit{OSDI'04: Sixth Symposium on Operating System Design and Implementation}, Dec 2004.
\bibitem[Hadoop(2007)]{ApacheHadoop} Apache Hadoop, open source MapReduce implementation, http://hadoop.apache.org/.
\bibitem[a(1900)]{a} Section 1, Par 1: Topics within astronomy require data gathered at the limits of detection of telescopes.  REF NOT FOUND YET.
\bibitem[SWarp(2002)]{SWarpA} Bertin, E. et al. 2002: The TERAPIX Pipeline, ASP Conference Series, Vol. 281, 2002 D.A. Bohlender, D. Durand, and T.H. Handley, eds., p. 228.
\bibitem[SWarp(2010)]{SWarpB} Bertin, E., SWarp, http://www.astromatic.net/software/swarp.
\bibitem[SDSS DR7(2009)]{SDSSDR7} SDSS Dr7 http://www.sdss.org/dr7/start/aboutdr7.html
\bibitem[Montage(2010)]{Montage} Montage, http://montage.ipac.caltech.edu/docs/.
\bibitem[Montage(2009)]{Montage2} Berriman, B. and Good, J., Building Image Mosaics With The Montage Image Mosaic Engine, Greater IPAC Technology Symposium, 2009.
\bibitem[TeraGrid(2010)]{TeraGrid} TeraGrid, https://www.teragrid.org/.
\bibitem[White(2009)]{White} White, T.  \textit{Hadoop The Definitive Guide}.  O'Reilly Media Inc. 2009, p. 103.
\end{thebibliography}
\end{document}